\documentclass{emulateapj}
\usepackage{apjfonts}
\usepackage{graphicx}
\usepackage[dvips]{color}

\shorttitle{Small PAHs in the Red Rectangle}
\shortauthors{Vijh et al.}

\begin{document}

\title{Small PAHs in the Red Rectangle}

\author{Uma P. Vijh\altaffilmark{1}, Adolf N. Witt\altaffilmark{1}}
\affil{Ritter Astrophysical Research Center, University of Toledo, Toledo, OH 43606}
\email{uvijh@astro.utoledo.edu, awitt@dusty.astro.utoledo.edu}
\and
\author{Karl D. Gordon}
\affil{Steward Observatory, University of Arizona, Tucson, AZ 85721}
\email{kgordon@as.arizona.edu}

\altaffiltext{1}{Visiting Astronomer, Cerro Tololo Inter-American Observatory.
CTIO is operated by AURA, Inc.\ under contract to the National Science
Foundation.}

\begin{abstract}
Following our initial discovery of blue luminescence in the spectrum of the Red Rectangle (RR) and its identification as fluorescence by small three- to four-ringed polycyclic aromatic hydrocarbon (PAH) molecules, we report on the spatial correlation between the blue luminescence and the 3.3 $\micron$ emission, commonly attributed to small, neutral PAH molecules, and on the newly-derived UV/optical attenuation curve for the central source of the RR, HD 44179. Both results provide strong additional evidence for the presence of small PAH molecules with masses of less than 250 a.m.u. in the RR, which supports the attribution of the blue luminescence to fluorescence by the same molecules. We contrast the excellent spatial correlation of the two former emissions with the distinctly different spatial distribution of the extended red emission (ERE) and of the dust-scattered light within the RR. The UV/optical attenuation curve of the central star is unlike any interstellar extinction curve  and is interpreted as resulting from circumstellar opacity alone. Major contributions to this opacity are absorptions in broad bands in the mid-UV, contributing to the electronic excitation of the luminescing PAH molecules, and a sharp ionization discontinuity near 7.5 eV in the far-UV, which places a sharp upper limit on the masses of the PAH molecules that are responsible for this absorption. The strength of the far-UV absorption leads to an abundance of the PAH molecules of $10^{-5}$ relative to hydrogen in the RR. Such small PAHs are perhaps unique to the environment in the RR, where they are shielded from harsh radiation by the dense circmstellar disk.
\end{abstract}



\section{INTRODUCTION}
Recently, we reported the discovery of blue luminescence (BL)\citep{vijh04} in the spectrum of the bipolar, proto-planetary Red Rectangle (RR) nebula \citep{cohen75,cohen04}. The emission, peaking near 378 nm, is contained in a band of FWHM $\sim$ 45 nm. The band-integrated intensity of the BL is comparable to that of the scattered light intensity and of the intensity of the extended red emission (ERE), which is also exceptionally strong in this object \citep{schmidt80,wb90}. This suggests that the BL is produced by a carrier of substantial abundance in the gas/dust  medium in the RR.

Through a comparison of the observed spectrum with laboratory photoluminescence spectra of a collection of likely dust components and with gas-phase fluorescence spectra of polycyclic aromatic hydrocarbon (PAH) molecules, \citet{vijh04} identified neutral three- and four-ring PAH molecules  such as pyrene, anthracene and phenanthrene as the most likely sources of the BL. This identification also appears consistent with several other important facts. Firstly, the RR is the brightest  known source of emission in the unidentified infrared (UIR) features \citep{rsw78,geballe85}, which is commonly attributed to PAH molecules and ions \citep{bakes04} which suggests high relative abundances and optimal excitation conditions for such molecules. Also, as a post-AGB star \citep{van03}, the carbon-rich central star of the RR, HD 44179, is in an active mass-losing and dust-producing stage of its life which suggests that PAH molecules are currently condensing in the carbon-rich outflow \citep{keller87, cherchneff92a}. This includes, in particular, smaller PAH molecules that are able to exist in the relatively benign radiation environment of the RR, but which would not be viable after being ejected into the much harsher radiation field of interstellar space, where three- and four-ringed PAHs are not expected to survive \citep{jochims94,jbl99}. And finally, anthracene and pyrene have exceptionally high flourescence efficiencies, which would cause their spectra to dominate any fluorescence spectrum of a mixture of PAH molecules \citep{brech94}. If confirmed by further observations, the detection of three- and four-ringed PAH molecules would represent the detection of the largest specifically identified molecules so far observed outside the solar system.

In this paper we report new observations and analysis dealing with the spatial distribution of the BL in the RR nebula. We show that the spatial distribution of the BL is distinctly different from that of the scattered light distribution in the same region of the spectrum and also distinctly different from that of the ERE. Significantly, there is an exceptionally close correlation between the distribution of the BL and the distribution of the UIR-band emission at 3.3 $\micron$, attributed to the C--H stretch transition in neutral PAH molecules, which provides strong support for the identification of the BL as fluorescence from relatively small PAH molecules. In the second part of this paper we discuss the newly-derived attenuation curve for the central star, HD 44179. The attenuation is shown to be circumstellar rather than interstellar in nature and to be dominated by absorptions attributable to electronic bound-bound and bound-free transitions in small PAH molecules.

\section{OBSERVATIONS}
Low-resolution, long slit spectra of the RR were obtained with the R-C (Cassegrain) spectrograph at the Cerro Tololo Inter-American Observatory (CTIO) 1.5 m telescope. These observations were made on 2003 March 26 and 29, using a 2$\farcs$5 wide slit, 7$\farcm$7 long. For observations in the blue (340 - 600 nm), grating \#09 with 300 l mm$^{-1}$ was used which is blazed at 4000 \AA, provides a 8.6 \AA\ resolution and a spectral coverage of 260 nm. Using a CuSO$_4$ filter to select the grating's first order, the setup covered a wavelength range of 340-600 nm, including the range from H$_\beta$ to the Balmer discontinuity. Grating \#11 with 158 l mm$^{-1}$, blazed at 8000 \AA, with a 16.4 \AA\ resolution was used in the first order with a GG495 cut-on filter for the red observations with a wavelength coverage from 500-1000 nm. The Loral IK CCD detector yielded a spatial scale of 1$\farcs$3 pixel$^{-1}$ along the slit. A coronographic decker assembly was used to minimize scattered light from the star while probing as much of the inner nebula as possible. All observations were made using the full extent of the 7$\farcm$7 long slit to get simultaneous sky observations. Spectra were taken at two nebular locations, 2$\farcs$5 and 5$\arcsec$ south of the central star HD 44179 with the slit in E-W direction (PA = 90$\arcdeg$). The nebular exposures were bracketed by exposures of the central star. Individual exposures were limited to 5-10 minutes on the nebula and 1 minute for the star, and 4-5 exposures were obtained for each orientation. Data reductions were carried out with IRAF 2.12 EXPORT, and all spectra were flux calibrated via observation of standard stars.

The 3.3 $\mu$m data are reproduced from \citet{kerr99}. The spectroscopic observations were made at the United Kingdom Infrared Telescope (UKIRT) on 1995 December 11. The spectra were taken with a slit 90$\arcsec$ long, 1$\farcs$2 wide, placed 5$\arcsec$ south of the central star HD 44179 (PA = -85$\arcdeg$).

For the FUV spectral energy distribution (SED) of HD 44179 archival observations from the International Ultraviolet Explorer (IUE) LWP 22416, SWP 38188, and LWR 04273 were used.

David  Malin (David Malin Images) provided us with one of the few existing deep blue images of the RR for the comparison of our spatially resolved spectra with morphological details of the nebular structure. The image, reproduced in Figure 1,  was taken by David Malin with the AAT at the Anglo-Australian Observatory. The exposure time was 15 min, the detector was a Kodak IIa-O plate (blue sensitive) with a GG385 cut-on filter. The effective bandpass extends from 390 nm to 480 nm. A red image of the RR taken with the \textit{HST} has been reproduced to aid in the correlation of the observed features with the nebular structure.

\section{ANALYSIS and RESULTS}
\subsection{Blue Luminescence in the Red Rectangle}
\begin{figure}
\plotone{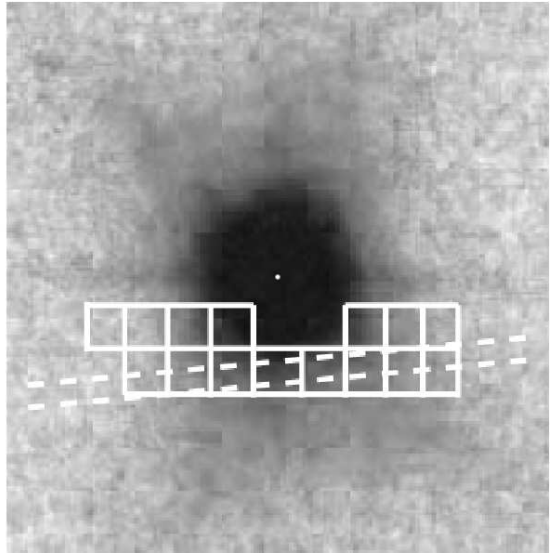}
\caption{Blue image of the Red Rectangle with overlaid extraction apertures. The size of the image is $30\arcsec \times 30\arcsec$. The white dot is an overlaid pixel to indicate the position of the central source. The dashed lines indicate the slit used for the 3.3 \micron\ observation. North is to the top, east to the left.\label{ble}}
\end{figure}

\begin{figure}
\plotone{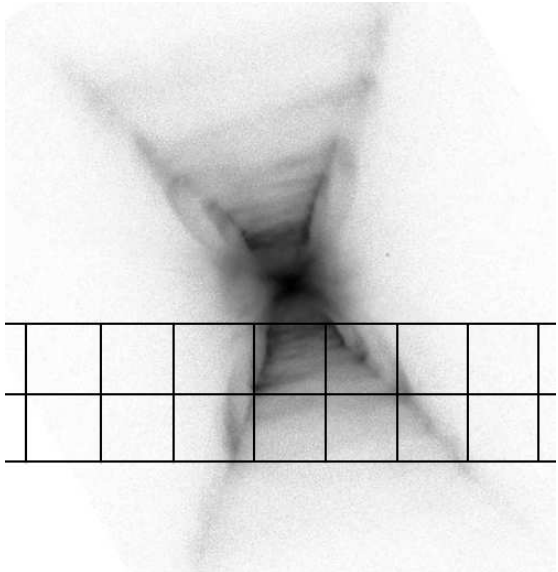}
\caption{\label{hst-red}Red image of the RR with overlaid extration apertures. The size of the image is $19\farcs4 \times 20\farcs2$. The \textit{HST} image (Credits: NASA, ESA, Hans Van Winckel (Catholic University of 
 Leuven, Belgium), and Martin Cohen (University of California, Berkeley)) has been reproduced to aid the identification of the nebular structure with the extracted apertures. North is to the top, east to the left.}
\end{figure}

Nebular spectra were extracted in two-pixel bins along the two slits with 7 apertures along the 2$\farcs$5 south slit and 8 apertures along the 5$\arcsec$ south slit with a spatial scale of 2$\farcs$6 per aperture. Thus, each spectrum represents a 2$\farcs$5 $\times$ 2$\farcs$6 region of the nebula. Figure \ref{ble} shows the blue image of the nebula overlaid with the BL apertures and the 3.3 $\micron$ slit and Figure \ref{hst-red} is a red image overlaid with the central ERE apertures. The line-depth technique \citep{vijh04,wv04} was used to detect and measure the BL at the positions of the hydrogen Balmer lines for each spectrum. Measurement of the Balmer discontinuity in the nebular spectra compared to that in the stellar spectrum is used to infer the BL at the Balmer discontinuity (see Appendix A for details). Measurement of the equivalent widths of the absorption lines could also be used for a similar analysis, however a small uncertainty in the determination of the continuuium levels over the steep spectral shape results in a relatively large uncertainity in the equivalent width. In contrast the errors in the line-depth measurements are essentially limited to the photon noise in the spectrum. The co-added nebular spectra had a signal-to-noise ratio per pixel of $\sim$ 10 in the outermost regions to $\sim$ 250 closer to the central star in the blue and a range of $\sim$ 20 - 225 in the red region of the spectrum. The stellar spectra had a signal-to-noise ratio per pixel of $\sim$ 500 in both spectral ranges.


\subsubsection{Spectral Variability in the BL}
\begin{figure}
\includegraphics[angle=270,width=3.3in]{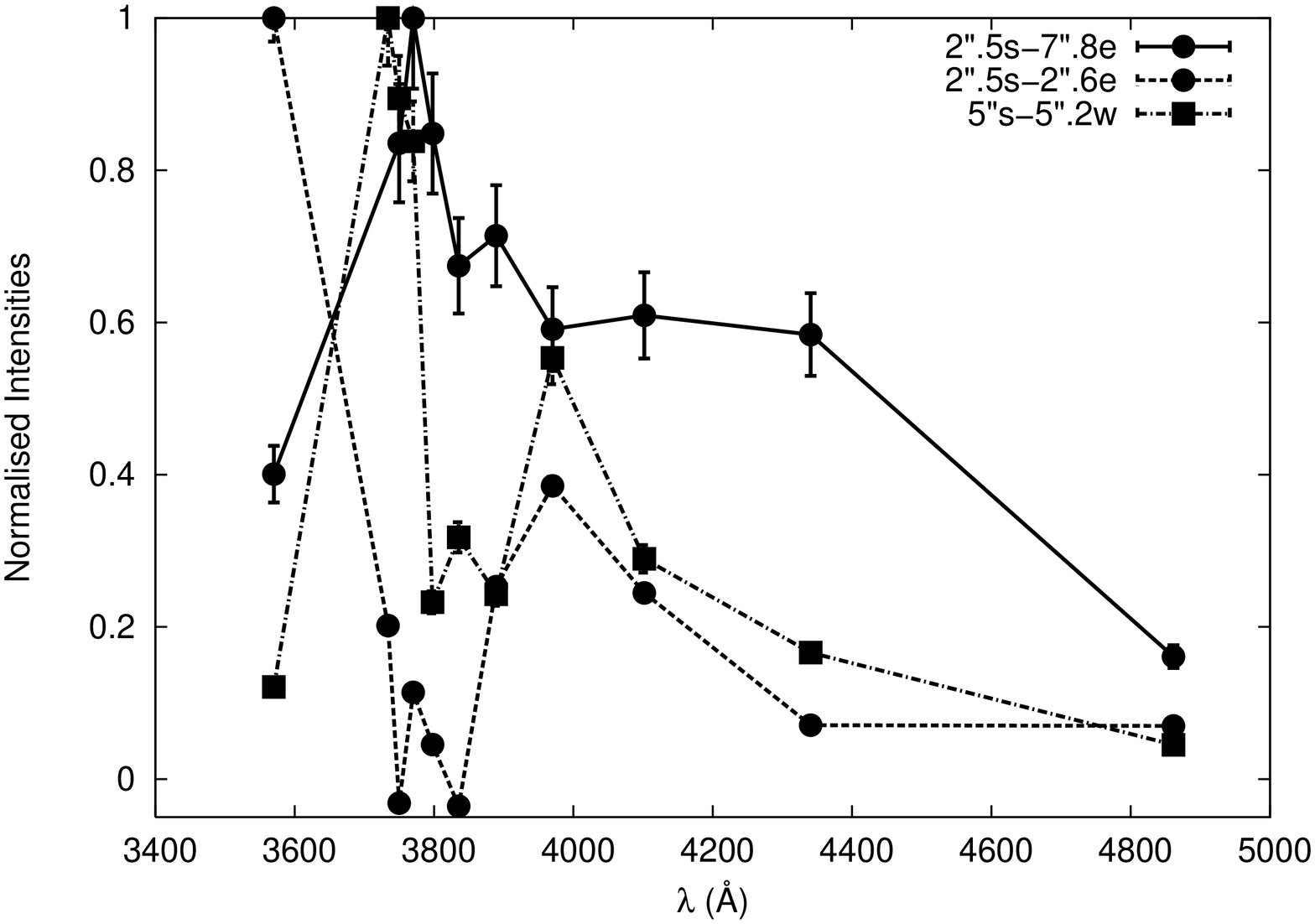}
\caption{Spectral variation of the BL. The solid line traces the intensity at 2$\farcs$5 south, 7$\farcs$8 east, the dashed line at  2$\farcs$5 south, 2$\farcs$6 east and the dash-dotted line at  5$\arcsec$ south, 5$\farcs$2 west positions. Solid circles indicate wavelength positions associated with the 2$\farcs$5 south slit and solid squares indicate wavelength positions associated with the 5$\arcsec$ south slit. Intensities are normalized to unity at the peak positions.\label{spectral-var}}
\end{figure}

The BL spectrum is not identical in different regions of the nebula and certain trends can be noted. The primary peak in the spectrum is near 378 nm \citep[Fig. 4]{vijh04}. A secondary peak around 397 nm starts to develop as we probe regions farther from the central star, more prominently along the 5$\arcsec$ south slit. Also, farther from the star, the BL intensities at longer wavelengths start to become stronger compared to the peak intensity. Interestingly, spectra that probe regions inside the conical outflow indicate that the BL in these regions has another peak at much shorter wavelengths, shortward of 360 nm. Figure \ref{spectral-var} depicts the BL spectrum at three such representative locations: 2$\farcs$5 south, 7$\farcs$8 east, a position in the outer regions of the nebula; 2$\farcs$5 south, 2$\farcs$6 east, a region close to the central source, inside the bipolar outflow; 5$\arcsec$ south, 5$\farcs$2 west, a region along the farther slit just outside the cone wall. 

Based on the existing correlation between the wavelength of peak intensity of the fluorescence spectrum and the molecular size of PAH molecules \citep{vijh04}, these trends can be interpreted to suggest a size variation in the BL carrier at different locations in the nebula: closer to the central source we see an emergence of a smaller population of PAHs and farther out we see increasing contributions from larger emitters.  Given the limited spectral resolution inherent in our detection technique and only fair spatial resolution these trends are at best an indication of the size distribution of the BL emitters. We defer a more complete analysis of the spectral variation of the BL spectrum until after completion of an ongoing program in collaboration with D.G. York which maps the complete RR nebula with dense spatial coverage using the 3.5 m telescope at Apache Point Observatory.


\subsubsection{Gradients of BL/Scattered Light}
\begin{figure*}
\includegraphics[angle=270,width=3.5in]{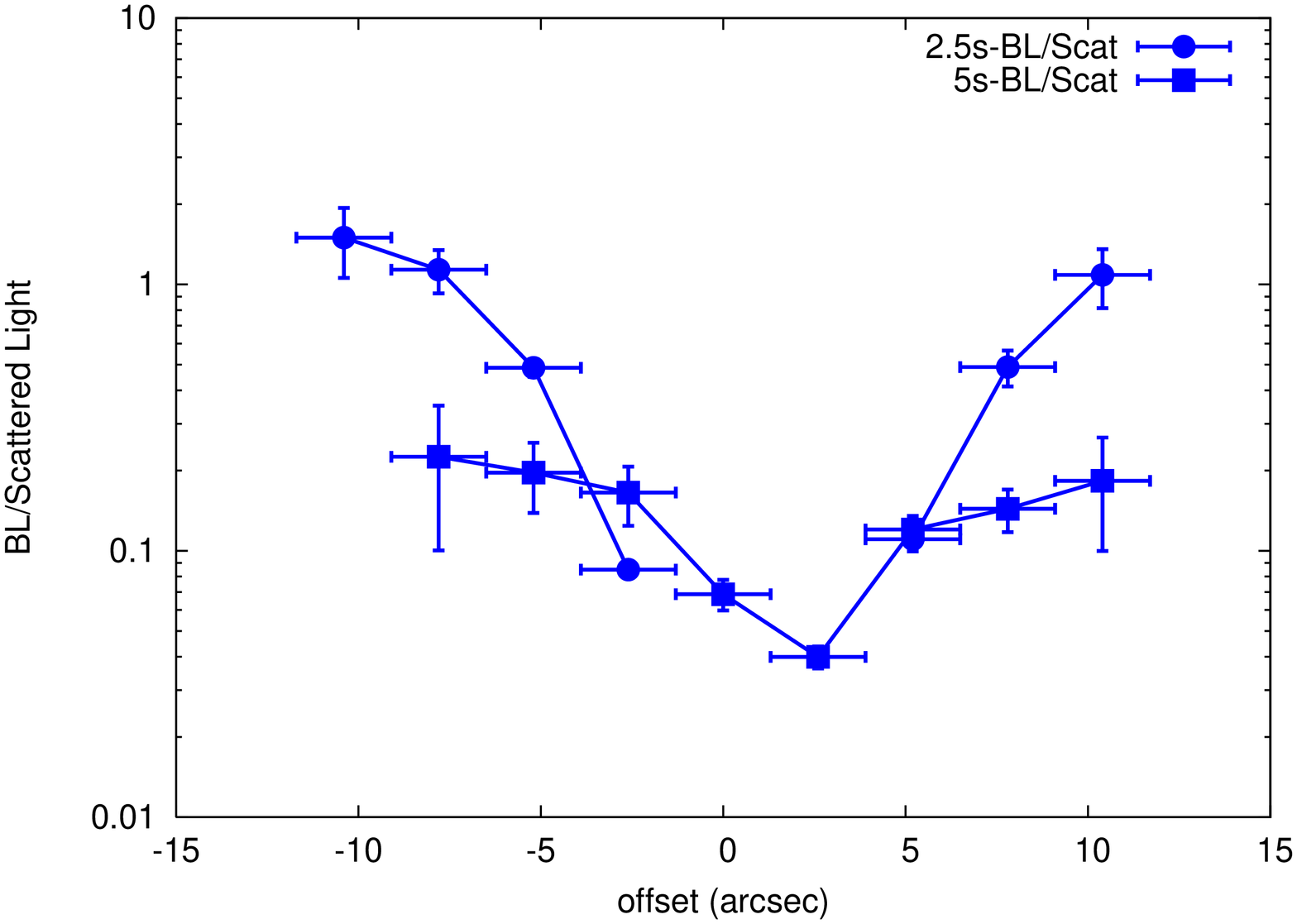}\includegraphics[angle=270,width=3.5in]{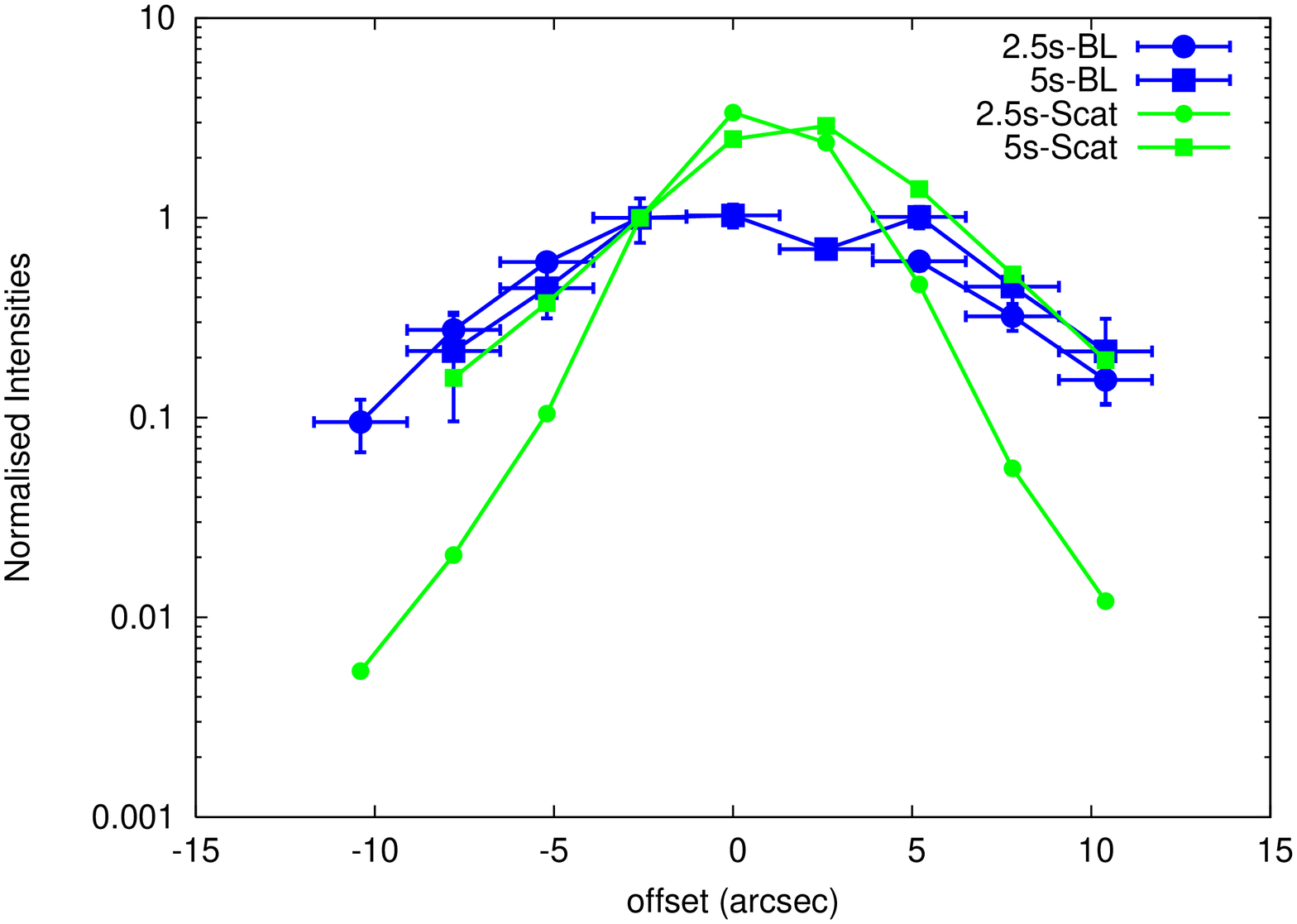}
\caption{a. Ratio of band-integrated BL to scattered light along the two slits. Solid line traces the ratio along the 2$\farcs$5 south slit and the dashed line that along the 5$\arcsec$ south slit. b. Normalized BL and scattered light intensities. The BL is traced by solid lines and the scattered light by dashed lines. Intensities are normalized to unity at 2$\farcs$6 east position. In both figures, solid circles denote spatial positions along the 2$\farcs$5 south slit and the solid squares those along the 5$\arcsec$ south slit. Positive and negative offsets indicate positions to the west and east respectively, along the slit. Zero offset refers to positions directly south of HD 44179.\label{blratio}}
\end{figure*}

 Figure \ref{blratio}a shows the distribution of the ratio of band-integrated BL to the total scattered light in the same band at various positions along the two slits. Figure \ref{blratio}b illustrates the spatial distribution of the BL and the scattered light intensities on a normalized scale. Due to the anisotropy of the phase function, scattered light is strongly forward directed and falls off steeply as we probe regions where our line of sight penetrates regions of the nebula where larger scattering angles predominate. The BL, on the other hand is due to fluorescence and is an isotropic emission and thus falls off less steeply. Thus, the spatial variation of the BL and scattered light along the two slits reveals the isotropic nature of the BL. Therefore, the ratio I$_\mathrm{BL}$/I$_\mathrm{Scat}$ increases away from the star, whereas the actual intensities fall away from the star as the density of emitters and the exciting radiation both decrease. It is equally illuminating to note that the BL intensities have similar profiles along the two slits, whereas the distribution of the scattered light it quite different; along the slit closer to the star, the scattered light falls off more steeply than along the second slit as the relative change in the distance from the central star varies more slowly in the latter case. 


\subsubsection{Correlations with Other Emissions}
\begin{figure*}
\includegraphics[angle=270,width=3.5in]{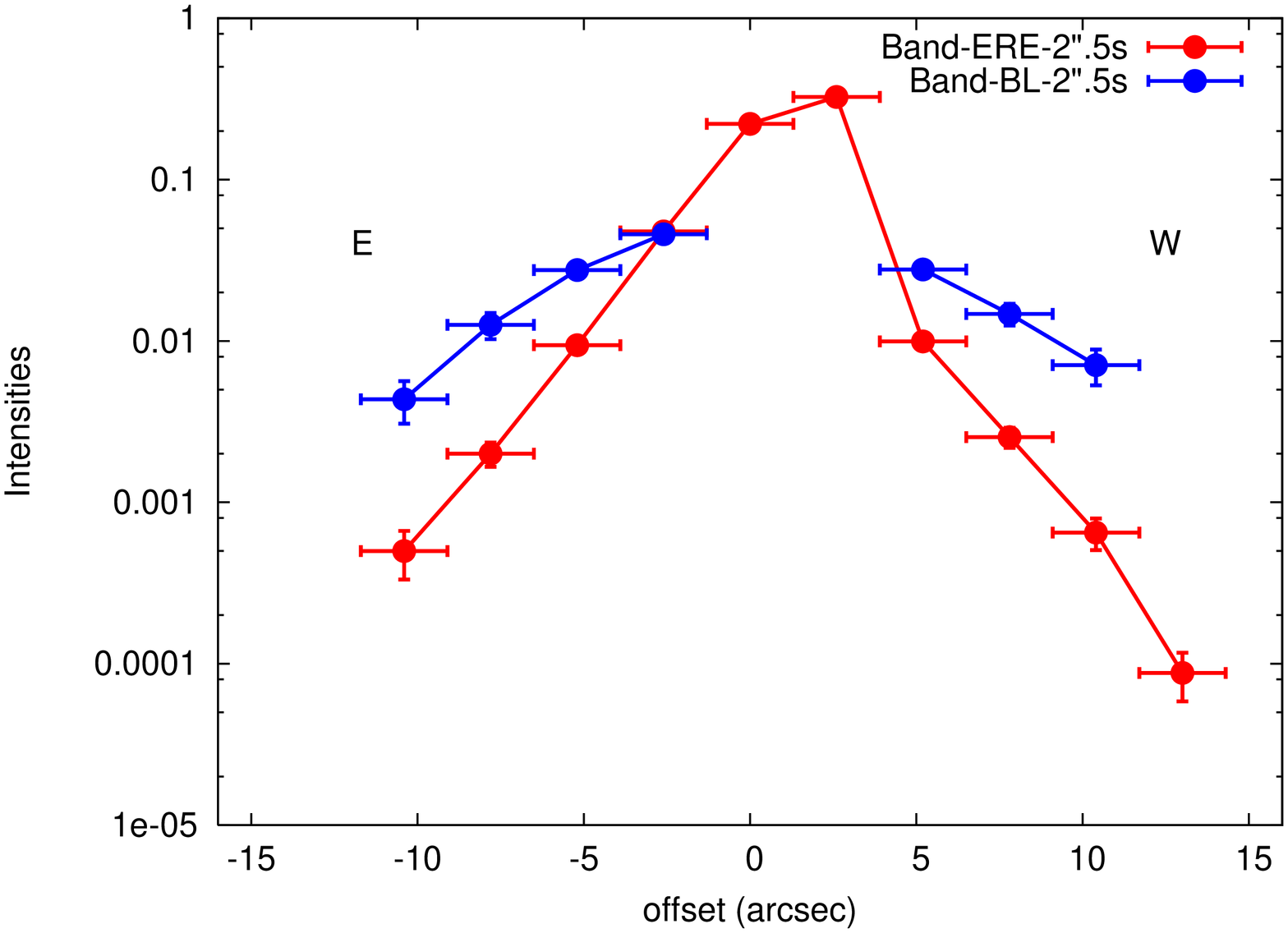}\includegraphics[angle=270,width=3.5in]{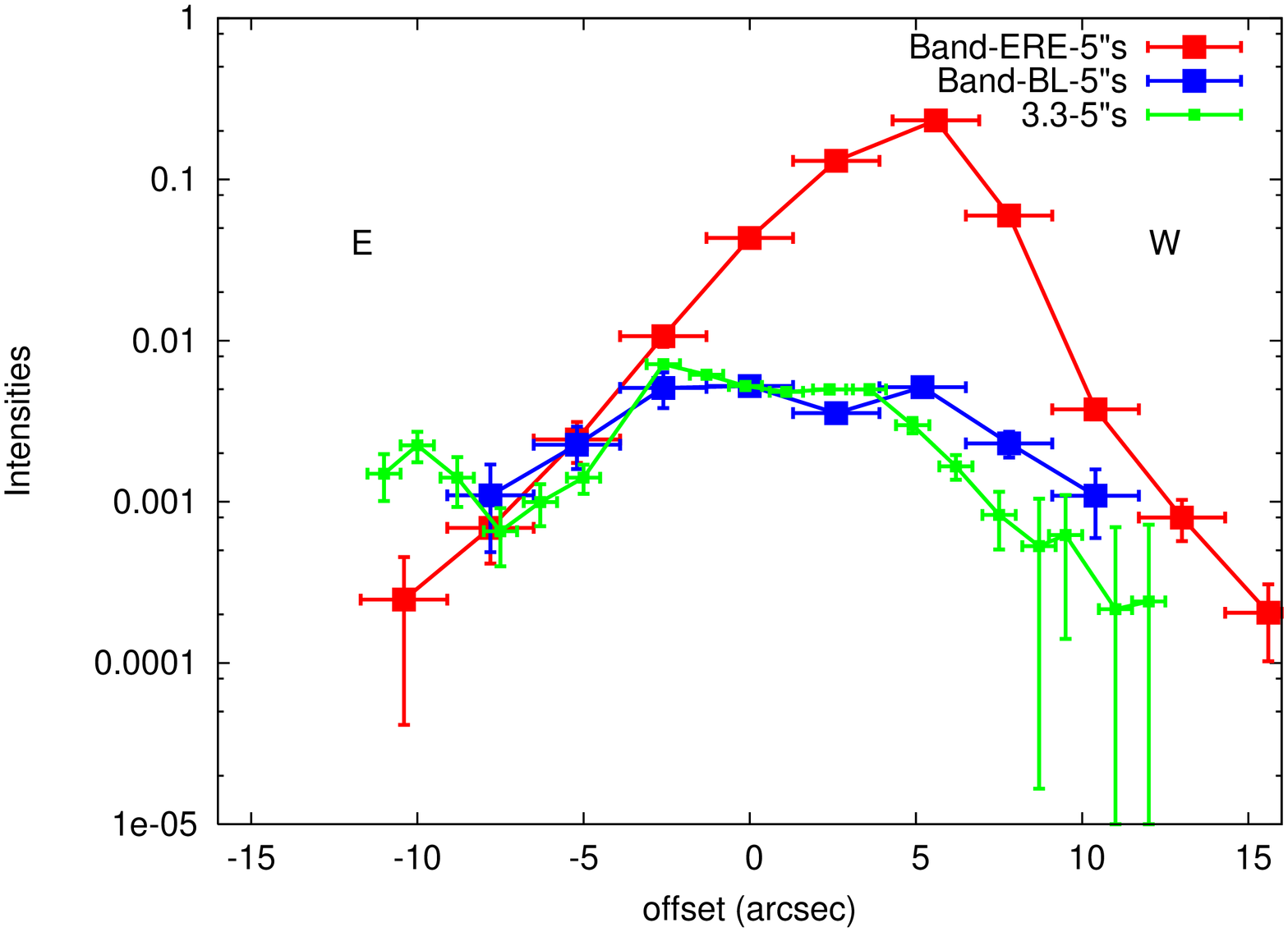}
\caption{\label{ere-bl-3.3}Spatial distribution of the BL and correlations with other emissions. Band-integrated intensities of BL and ERE at positions along the two slits (a. along the $2\farcs5$ south slit and b. along the $5\arcsec$ south slit) are plotted. ERE is plotted using dashed lines, the BL using solid lines and the 3.3 $\micron$ emission normalized to the value of the BL at 5$\arcsec$ south, zero offset position, is plotted with a dotted line. Filled circles indicate positions along the 2$\farcs$5 south slit and filled squares those along the 5$\arcsec$ south slit. The intensities are band-integrated and are given in units of ergs cm$^{-2}$ s$^{-1}$sr$^{-1}$ for the BL and the ERE.}
\end{figure*}

Figure \ref{ere-bl-3.3} displays the spatial distribution of the band-integrated intensities of ERE and the BL along the two slits and the 3.3 $\micron$ UIR-band emission along a 5$\arcsec$ south (PA=-85$\arcdeg$) slit (Fig. \ref{ble} and \ref{hst-red} show the apertures). The 3.3 $\micron$ intensites (obtained from \citet{kerr99}) have been normalized to the intensity of the BL at the 0 east-west, 5\arcsec south position. This figure reveals a number of interesting facts. The 3.3 $\micron$ UIR-band emission and the BL are almost perfectly correlated, which we will discuss in more detail in the next paragraph. The spatial distribution of the BL is distinctly different from that of the ERE, the ERE is strongly peaked close to the central source and on the X-shaped walls of the outflow cone while the BL has a broader, more diffuse distribution.

The 3.3 $\micron$ emission correlates exceptionally well with the BL along the 5$\arcsec$ south slit, and the small differences are probably attributable to the slightly different widths and PA of the two slits (see Fig. \ref{ble}). The 3.3 $\micron$ emission feature belongs to the distinctive set of mid-infrared emission features which have been attributed to emission from PAH molecules, which become vibrationally excited upon absorption of a UV photon and subsequently relax with the emission of IR photons \citep{bakes04,schutte93}, in addition to electronic relaxation through fluorescence or phosphorescence transitions. Most of the vibrational transitions are insensitive to molecular sizes, but the 3.3 $\micron$ emission attributed to C--H stretch from PAH molecules is sensitive to the size of the molecular species. PAHs having $\sim$ 20 carbon atoms show 3.3 $\micron$ emission intensities $\sim$ 50 times stronger than that from PAHs having $\sim$ 100 carbon atoms \citep{schutte93}. Earlier observations of the RR showed a pronounced lack of similarity between the 3.3 $\micron$ and the 11.3 $\micron$ surface brightness distributions \citep{breg93}, the latter being attributed to C--H out-of-plane bending modes in relatively large PAH molecules. Furthermore, the 3.3 $\micron$ emission traces neutral PAH species \citep{bakes04}, and becomes extremely weak in ionized PAHs. The fact that the BL follows the 3.3 $\micron$ distribution so closely indicates that the BL emitters and the 3.3 $\micron$ emitters are similar small, neutral PAHs and thus strengthens our identification of the BL as fluorescence by small neutral PAHs.

The ERE, BL and the 3.3 $\micron$ emission are all isotropic emissions and yet the spatial distributions of the ERE is dissimilar to those of the BL and the 3.3 $\micron$ emission. The  ERE falls off much more steeply than the other two that have a broader distribution. The ERE is largely confined to the bipolar, X-shaped structure of the nebula \citep[see also][Fig. 2,]{sw91}, whereas the BL is the dominant emission in the outer regions. The variation in the spatial distributions of these different emissions must be attributed the distribution of the carriers and the respective exciting radiations. The BL carriers, most likely small PAHs, are probably ionized inside the bipolar ouflow cone, where they are exposed to direct illumination from the central RR source. Thus fluorescence in these regions is effectively quenched. Farther out, outside the cones, the molecules/carriers are shielded from the far-UV ionizing radiation but still receive through scattering the mid-UV exciting radiation and thus emit with fairly high intensities. The ERE carriers on the other hand seem more robust as the ERE intensities are high closer to the star inside the bipolar outflow cone and along the walls (X-shaped structure). This could, in fact be interpreted as pointing toward some unknown \emph{ionized} species as the carrier of the ERE. Outside the outflow region the ERE intensities drop by over three orders of magnitude, suggesting that either the carriers of the ERE do not exist in such regions or if they do exist they do not receive the exciting radiation needed to produce the ERE.


\subsection{ATTENUATION OF HD 44179}

\subsubsection{\label{sec:attn}Visual Reddening and Attenuation}
The central source of the RR, HD 44179, is a single-line spectroscopic binary star \citep{waelkens92, waelkens96, vww95, menshchikov02} that is totally hidden from direct view at optical wavelengths by a circum-binary disk seen nearly edge on. Diffraction-limited imaging \citep{cohen04} reveals the central source  as a pair of scattered-light lobes above and below the disk, directing light from the hidden central stars via scattering by dust toward the Earth. The photometrically dominant component of the binary was classified as spectral type B9/A0 \citep{cohen75} originally, but is now recognized as a post-AGB star with an effective temperature near 8000 K with a highly metal-deficient atmosphere typical of post-AGB stars \citep{van03, menshchikov02}. The low metal abundance, e.g. [Fe/H] = -3, gives this star the spectroscopic appearance of a much hotter B9/A0 star, but with an intrinsic (\bv) color characterized by its actual effective temperature. By convolving the appropriate Kurucz atmosphere model (T$_{\mathrm{eff}}$ = 8250 K, log g = 1.5, [Fe/H] = -3.0) with the B and V pass bands, we determine an intrinsic (\bv) = -0.04 mag for the photometrically dominant component of HD 44179. We note that the nature of the secondary component is not known directly from observation, but its inferred mass of 0.35 M$\sun$ and the fact that the innermost region of the RR contains a small H \textsc{ii} region suggest that it is most likely a hot white dwarf with a lumininosity of less than 2 \% of that of the primary \citep{menshchikov02}.

The observed (\bv) color of HD 44179 has been inferred from broadband photometry of the RR by \citet{cohen75} as 0.39 mag., leading to an estimated color excess of E(B$-$V) = 0.35 mag. The peculiar nature of HD 44179, its location within an optically thick disk, and its uncertain distance make it difficult to estimate the total visual attenuation of its light as seen from Earth. Here we adopt the latest model \citep{menshchikov02} which assigns a distance of 710 pc and a total luminosity of 6050 L$_\sun$ to the central object, with uncertainties of 10 \% and 20 \%, respectively. Given a visual brightness of V = 8.83 \citep{cohen75}, we then estimate a visual attenuation of A$_\textrm{V}$ = 4.2 mag. Here, we deliberately distinguish between attenuation and extinction. By extinction we understand the partial obscuration of \textit{direct} star light due to absorption and scattering by dust along the direct line-of-sight to the star, while attenuation measures the reduction of observable flux due to absorption and scattering in a complex geometry, where the residual observable flux may indeed only be scattered light re-directed into the line of sight while the direct light is totally absorbed, as is the case in the RR.

The attenuation of the light of HD 44179 is characterized by a ratio R$_\mathrm{V}$ = A$_\mathrm{V}$/E(\bv) = 11.9, much larger than the average value of R$_\mathrm{V} \sim$ 3.1 encountered with interstellar extinction. The attenuation in the UV/visible portion of the spectrum of HD 44179 is therefore much greyer than typical interstellar extinction, an effect attributed in the past to extinction by larger grains in the RR \citep{menshchikov02}. We consider it more likely that the large value of R$_\mathrm{V}$ arises from the fact that only scattered light from HD 44179 can be observed at optical wavelengths. Upon first scattering, scattered light is initially bluer than that of the illumination source, but upon transfer through an optically thick medium and multiple scattering, scattered light also reddens but at a slower rate than expected from the optical path length covered, as shown in the example of the colors of reflection nebulae by Witt (1985). The high value of the estimated visual attenuation of A$_\mathrm{V}$ = 4.2 mag and the fact that scattered light must be re-directed by 90 degrees require that multiple scattering is the dominant process for the transfer of optical radiation from the center of the RR.

\subsubsection{UV/Optical Attenuation Curve}
\begin{figure}
\includegraphics[height=5.0in]{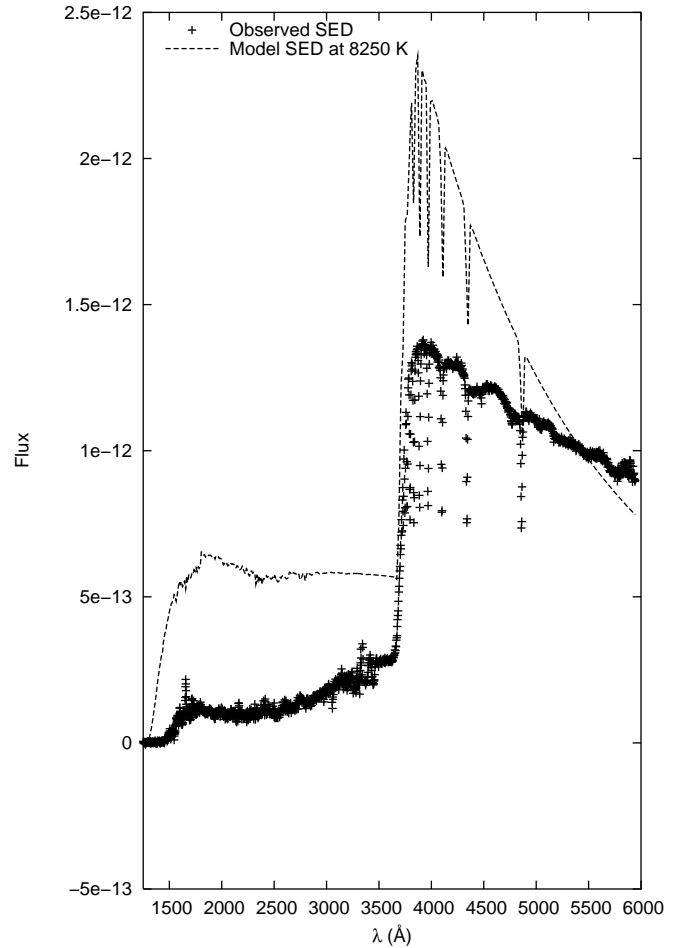}
\caption{\label{sed}Observed UV/optical SED of HD 44179 compared to model SED. The observed SED has been obtained by combining the archived flux distributions in the near- and far-UV from IUE and optical flux distribution obtained by our CTIO spectroscopic measurements. The model SED is from a Kurucz (1993) stellar atmosphere for [Fe/H]=-3, log g =1.5, and T$_{\mathrm{eff}}$=8250 K, normalized to the obsereved SED at the V-band. The stellar fluxes are displayed in units of ergs cm$^{-1}$s$^{-1}$\AA$^{-1}$.}
\end{figure}

In Figure \ref{sed} we display the observed (attenuated) UV/optical SED of HD 44179 over the wavelength interval from 130 nm to 600 nm. We produced this SED by combining the observed flux distributions in the near- and far-UV obtained from the IUE archive (LWP 22416, SWP 38188, and LWR 04273) with our optical SED obtained from CTIO spectroscopy. No adjustments or corrections were applied; the continuity between the ground-based and space-based observations is almost perfect, indicating that the two sets of absolute calibrations are in excellent accord. Also shown in Figure \ref{sed} is a model SED for a metal-deficient, giant-star using a \citet{kurucz93} stellar atmosphere for [Fe/H] = -3 , log g = 1.5 and T$_\textrm{eff}$ = 8250 K, which we propose as a suitable representation of the intrinsic SED for the photometrically dominant component of HD 44179. The model SED is normalized to the observed SED at V. The effective temperature of this model atmosphere is higher than that proposed by \citet{menshchikov02} by 500 K, but our choice was dictated by the requirement that the model and the observed SED should have the same relative strength of the Balmer discontinuity such that the ratio of these two SEDs would be continuous across the Balmer discontinuity. Model SEDs with temperatures differing from 8250 K by as little as 250 K exhibit noticeable discontinuities, when divided into the observed SED.


\begin{figure}
\includegraphics[height=5.0in]{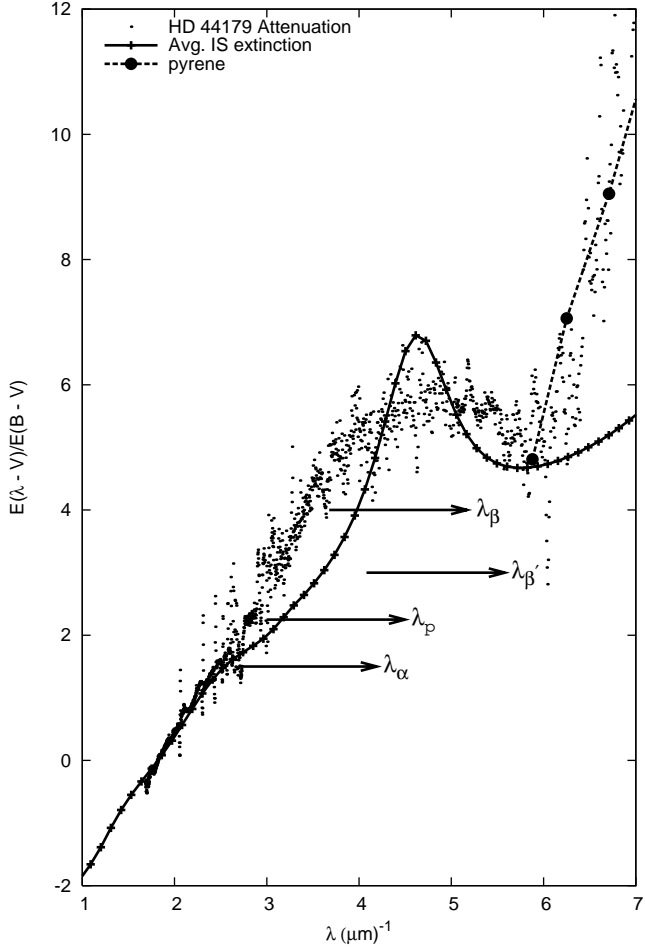}
\caption{\label{attn}UV/optical attenuation curve for HD 44179. The average interstellar extinction curve is shown for comparison with the solid line. Also shown is the expected contribution of PAH ionization to the far-UV part of the attenuation curve by pyrene as a representative example with a dashed line. Indicated schematically are the four characteristic PAH absorption bands, the $\alpha$, para, $\beta$ and the $\beta^\prime$. The curves are normalized to E(\bv)=1.}
\end{figure}

Figure \ref{attn} displays the wavelength dependence of the UV/optical attenuation suffered by the light of HD 44179. The attenuation curve is normalized to E(\bv) = 1, and it is shown in comparison with the average Galactic interstellar extinction curve with identical normalization. Several distinct differences between the two curves are immediately apparent.  With the slopes forced to coincide in the \bv range, the observed attenuation curve rises more steeply in the near-UV compared to the interstellar extinction curve. There is no sign of the 217.5 nm absorption peak; instead we see a very much broader attenuation hump reaching its peak near $\lambda^{-1}\sim$ 5 $\micron^{-1}$. Finally, near $\lambda^{-1}\approx$ 6.0 $\micron^{-1}$ the attenuation curve exhibits a sharp discontinuous rise, which is much more abrupt than the gentle rise shown by the extinction curve. 


\begin{figure}
\includegraphics[height=5.0in]{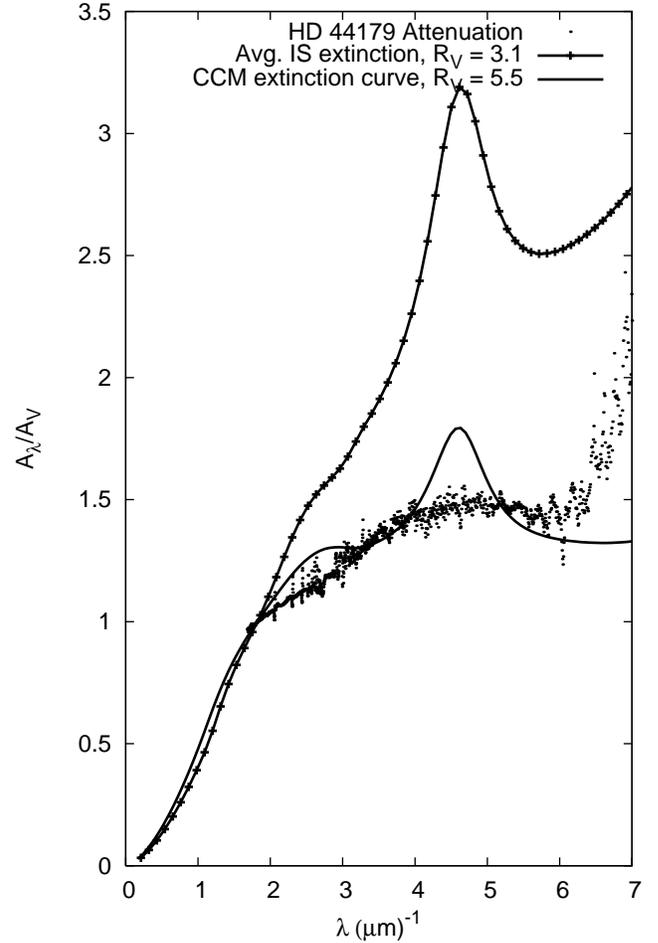}
\caption{\label{extn_av} The UV/optical attenuation curve for HD~44179, normalized to A$_\textrm{V}$ = 1 mag. For comparison, two Galactic extinction curves for R$_\textrm{V}$ = 3.1 and R$_\textrm{V}$ = 5.5 are shown as well. }
\end{figure}

As is well known, the perception of anomalous attenuation/extinction curves is strongly dependent upon their normalization \citep{fitzpatrick04}. In Figure \ref{extn_av}, therefore, we display the attenuation curve for HD 44179 normalized at V, scaled to A$_\mathrm{V}$ = 1.0, and compare it to Galactic extinction curves with R$_\mathrm{V}$ = 3.1, representative of the diffuse ISM, and R$_\mathrm{V}$ = 5.5, representative of a molecular cloud environment and a grain size distribution dominated by larger grains. While the overall level of UV attenuation in HD 44179 relative to the attenuation at longer wavelengths is roughly comparable to the R$_\mathrm{V}$ = 5.5 extinction curve and consistent with earlier findings of large dust grains in the RR \citep{menshchikov98,jura97}, the two features that make the HD 44179 attenuation truly unique remain clearly visible: the absence of the 217.5 nm absorption peak and the sharp discontinuous rise in attenuation at $\lambda^{-1}\approx$ 6.0 $\micron^{-1}$.


In two subsequent sections we will interpret the features just described in more detail. Given the overall appearance of the attenuation curve for HD 44179, we can conclude, however, that the suggestion by \citet{menshchikov02} that most, if not all, of the reddening of HD 44179 is caused by interstellar instead of nebular material intrinsic to the RR is almost certainly incorrect. The hydrogen column density (N(H I) $\leq 1\times10^{20}$ cm$^{-2}$) implied by the equivalent width of the 578 nm diffuse interstellar band in the spectrum of HD 44179 \citep{hobbs04} translates to an extinction of A$_\mathrm{V}<$ 0.05 mag. Also noteworthy is the absence of narrow, interstellar absorption lines Na I and Ca II in the spectrum of HD 44179 \citep{hobbs04}. The probability that the line-of-sight to one of the most unusual objects in the Galaxy should also be characterized by one of the most unusual interstellar extinction curves ever observed is vanishingly small. Given the large internal optical depth of the RR implied by the visual attenuation found in Sect. \ref{sec:attn}, we suggest that the unique characteristics of the attenuation curve of HD 44179 must be explained in terms of the optical characteristics of the molecular and dust constituents of the RR.

\subsubsection{Far-UV Rise}
\begin{figure}
\includegraphics[height=5.0in]{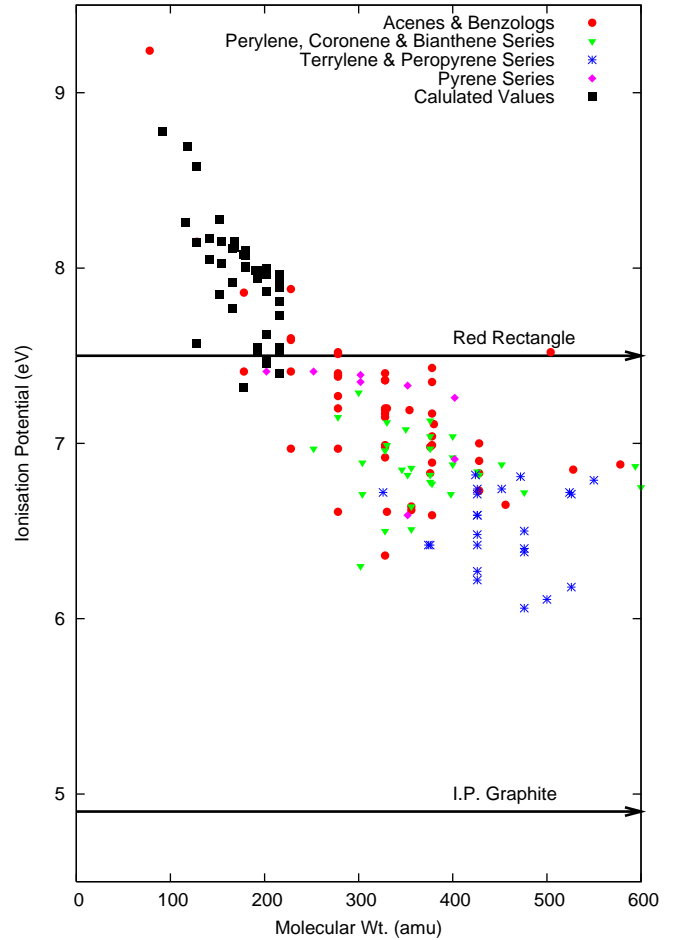}
\caption{\label{pahip}Ionization potential of different families of PAHs as a function of molecular mass. Also shown for reference are the IP of graphite \citep{cs75,burns69} and the ionization discontinuity apparent in the attenuation curve of the central source of the RR.}
\end{figure}

The abrupt and discontinuous rise in the observed attenuation curve beginning at $\lambda^{-1}\approx$ 6 $\micron^{-1}$, corresponding to a wavelength of 167 nm or a photon energy of 7.4 eV, is not characteristic of dust extinction but rather suggests the onset of strong absorption, e.g. the absorption associated with the ionization discontinuity in small PAH molecules, which occurs at just this energy. \citet{verstraete90} published detailed ionization cross-sections for two PAH molecules, pyrene and coronene, which exhibit a sharp onset at $\lambda^{-1}\approx$ 6.0 $\micron^{-1}$, similar to the observed rise in the HD 44179 attenuation curve. When normalized to cross-section per C-atom, the absolute values of the ionization cross-sections and their wavelength dependencies are almost indistinguishable. A typical value is $1.35\times10^{-18}\textrm{ cm}^2\textrm{ C-atom}^{-1}$ at 143 nm or $\lambda^{-1}\approx$ 7.0 $\micron^{-1}$. The work by \citet{jbl96} shows similar ionization cross-section spectra for naphthalene, azulene, anthracene, phenanthrene, and benz(a)anthrathene, all with an sharp onset near 7.5 eV, corresponding to $\lambda^{-1}\approx$ 6.07 $\micron^{-1}$. However, the ionization potential (IP) of PAH molecules is generally size-dependent in a manner similar to the size-dependence of the peak wavelength of fluorescence \citep{vijh04}. We demonstrate this dependence of the IP on molecular size by plotting known IP values \citep{es81} against their molecular mass in Figure \ref{pahip}. This graph illustrates that the rise in the far-UV attenuation curve is consistent with the ionization of PAHs, provided their molecular mass is not greater than about 250 a.m.u. A similar upper limit of the masses and sizes of PAH molecules was also deduced by \citet{vijh04} on the basis of the observed blue fluorescence seen in the RR. The attenuation curve suggests that absorption by molecules of this size and smaller could contribute to the sudden far-UV rise but that molecules of larger size cannot be present with significant abundance close to the central star; if they were, the discontinuity would occur at longer wavelengths.


An important question is whether the observed increase in attenuation in the RR over the wavelength interval from $\lambda^{-1}\approx$ 6.0 $\micron^{-1}$ to $\lambda^{-1}\approx$ 7.0 $\micron^{-1}$ is in agreement with the likely carbon column density and the measured ionization cross-section at $\lambda^{-1}\approx$ 7.0 $\micron^{-1}$. The observed increase in attenuation over this interval is 2.0 mag , when reduced to the actual color excess of E(\bv) = 0.35 of HD 44179, corresponding to an optical depth increase by $\Delta \tau$ = 1.85. With an ionization cross-section $\sigma = 1.35\times10^{-18}\mathrm{ cm}^2\ \mathrm{C-atom}^{-1}$ at 143 nm wavelength, the corresponding carbon column density is $\mathrm{N}_\mathrm{C} = 1.37\times10^{18}\mathrm{ cm}^{-2}$ for carbon atoms tied up in small PAH molecules with masses less than 250 a.m.u. If the visual attenuation of 4.2 mag is associated with a mixture of gas and dust with relative elemental abundances as found in the Sun, then N(H\textsc{ i} + 2H$_2$)/A$_\mathrm{V}$ = $1.87\times10^{21}\mathrm{ cm}^{-2} \mathrm{mag}^{-1}$ \citep{bsd78}, and N(H\textsc{ i} + 2H$_2$) = $7.85\times10^{21}\mathrm{ cm}^{-2}$. With a solar carbon abundance ratio of N$_\mathrm{C}$/N$_\mathrm{H}$ = $3.4\times10^{-4}$, this leads to a total expected column density of carbon in all forms of N$_\mathrm{C}$ = $2.7\times10^{18}\mathrm{ cm}^{-2}$. Hence, in order to produce the observed far-UV rise by ionization of small PAHs, we require that about 50 \% of all carbon atoms reside in small PAHs, if normal solar abundances prevail. Thus, the strength as well as the wavelength dependence of the observed far-UV rise in the attenuation curve are fully consistent with an origin in the ionization of small PAH molecules. In addition, HD 44179 is recognized as a carbon-rich post-AGB star whose relative carbon abundance must exceed that of the Sun by factors of two to six \citep{cherchneff92}, making the case even stronger. In Figure \ref{attn} we have indicated the expected contribution of PAH ionization to the far-UV part of the attenuation curve for pyrene as an representative example.  The fact that the observed rise is slightly steeper than the individual curve of a specific PAH molecule can be understood easily, if still smaller PAH molecules with higher IP values contribute to the absorption at the shorter wavelengths.

\subsubsection{Mid-UV Hump}
Given the facts that blue fluorescence from PAH molecules with an intensity comparable to that of the ERE contributes strongly to the the nebular emission of the RR and that the ionization of these molecules in the inner part of the RR contributes strongly to the far-UV attenuation at $\lambda^{-1}\gtrsim$ 6 $\micron^{-1}$, it appears most likely to us that the mid-UV hump in the attenuation curve is produced by absorption related to electronic bound-bound transitions in these same molecules. With a sharp onset of the fluorescence near 360 nm, the absorption powering this emission must occur at shorter wavelengths. According to \citet{clar72}, the absorption spectra of PAH molecules consist of four characteristic bands, the $\alpha$, para, $\beta$, and $\beta^\prime$ bands, located at progressively shorter wavelengths in the mid-UV. They are schematically indicated in the mid-UV portion of Figure \ref{attn}, with arrows starting at the band edges and the height indicating the relative strengths of the bands \citep[See also][]{schmidt77}. The $\beta$-band which typically occurs at a wavelength 1/1.35 of that of the $\alpha$ band, is typically two orders of magnitude stronger than the $\alpha$-band, while the intermediate para-band is typically one order of magnitude stronger than the $\alpha$-band. The wavelength positions of these bands are closely related to the IPs of the corresponding molecules \citep{cs75, cs77, cs79}, so that a continuum of sizes of small PAH molecules with IP $>$ 7.4 eV can be expected to produce a broad, quasi-continuous absorption band as observed in the attenuation curve of HD 44179. \citet{verstraete92} produced a generic mid- and far-UV absorption spectrum for interstellar PAH molecules, which exhibits the mid-UV hump and the far-UV rise so clearly apparent in the attenuation curve of HD 44179.

\section{DISCUSSION}
\subsection{Identification of the BL Carrier}
When we first attributed the BL in the RR to fluorescence by three- and four-ringed neutral PAH molecules \citep{vijh04}, the identification of the likely carrier was tentative for several reasons. Our observational technique for separating the BL spectrum from the underlying scattered light spectrum produces measurements of the BL intensities only at the positions of the hydrogen Balmer lines, yielding data with very limited and uneven spectral resolution that is incapable of revealing fine spectral features needed for specific identifications of individual molecular species. In addition, the fluorescence spectra of several likely molecular  PAH species occupy nearly the same wavelength interval, contributing to the observed spectrum in proportion to their relative abundances and their fluorescence quantum yields. Finally, the gas-phase fluorescence spectra of PAHs vary as a function of temperature, with sub-bands losing their sharpness with increasing temperature \citep{chi01}. For these reasons, the identification of the carrier of the BL can only be made in part on the basis of spectral comparison and must rely largely upon other supporting evidence such as presented in this paper. 

There are prospects for improving the spectral resolution of the BL spectrum by using other techniques for separating the BL spectrum from the scattered light spectrum. One such technique is spectropolarimetry. It relies on the fact that the scattered nebular light is highly linearly polarized \citep{perkins81} while the BL is expected to be unpolarized. The presence of the BL in the composite spectrum of the RR would then be revealed by a proportional reduction in the linear polarization of the combined light. The only two published sets of spectropolarimetric observations at optical wavelengths of the RR nebula \citep{schmidt80,reese96} do indeed reveal a sudden reduction of the linear polarization shortward of 410 nm consistent with the onset of the BL band. Unfortunately, the existing data lack the spectral resolution and adequate spectral coverage at shorter wavelengths that would be necessary. A program of spectropolarimetric observations of the RR is being planned.

At this stage, the strongest supporting evidence for the correctness of our identification of the carrier of the BL are the close spatial correlation with the 3.3 $\micron$ C--H stretch emission that is highly specific to small, neutral PAH molecules, and the evidence obtained from the analysis of the attenuation curve of the central source that exhibits strong features identifiable with electronic-band absorptions and an ionization discontinuity specifically pointing toward PAH molecules with masses of 250 a.m.u. or less.

\subsection{Spectral Variability of the BL}
Our initial results showing evidence of systematic spectral variability of the BL with position within the bi-polar structure of the RR are intriguing but clearly need more observations with the highest possible spatial and spectral resolution. Such observations would have the potential of probing the variation in the dominant sizes of the PAH molecules as a function of distance from the central source and as a function of position within  and outside of the bi-polar outflow cones and within the shadow of the circum-source disk. Our initial results are consistent  with an increase in the size of the most strongly fluorescing PAH molecules with increasing distance from the central source, with a superimposed azimuthal dependence with respect to the outflow axis of  the RR. This could reflect the growth of the molecules with increasing distance, but it could also be a result of the different excitation requirements for the fluorescence of PAH molecules of different sizes. The wavelengths of the absorption bands of PAH molecules are related to their respective ionization potentials \citep{cs75,cs77,cs79} and occur at shorter wavelengths for smaller PAHs, and at longer wavelengths for larger PAHs. As a consequence, exciting radiation for the smallest PAHs may be lacking at increased distances as a result of the high opacity of the intervening nebular material.

\subsection{Spatial Variation of the BL}
The close spatial correlation between the BL and the 3.3 $\micron$ emission along two nearly identically placed slits, positioned 5\arcsec south of HD 44179 is remarkable in itself by supporting the interpretation that both emissions are produced by the same molecules. It is further remarkable in that both emissions show a distinctly different distribution than the distribution of the scattered light in the RR. The latter is produced by dust with a strongly forward-directed scattering phase function, which when coupled with an embedded source, produces a strongly peaked brightness distribution of the resulting scattered light. This is clearly apparent when viewing the blue image of the RR (Fig. \ref{ble}), which is dominated by a bright circular source of about 8$\arcsec$ diameter containing most of the scattered light. The PAH fluorescence, both electronic in the form of the BL as well as vibrational in the form of the 3.3 $\micron$ emission, by contrast is much less centrally peaked, consistent with an isotropic emission process for both. It is further remarkable that the BL brightness distribution does not reflect the crossing of the X-shaped arms or whiskers of the RR, which are so prominent at red wavelengths \citep{cohen04}. This indicates that the BL is neither associated with the interior of the outflow cones nor with the walls of the outflow cavity but rather is associated with a population of emitters residing in a shell surrounding the inner bi-polar structure. This would place them into the shadow of the circum-source disk and thus outside the reach of ionizing radiation, consistent with our attribution of the BL to neutral PAHs.

The difference between the spatial distributions of the BL and the ERE is striking. It has been known for some time that the ERE is strongly concentrated in the X-shaped arms of the RR \citep{sw91}, which are thought to be the projections of the walls of the bi-polar outflow cavity upon the plane of the sky. Our data leave little doubt that the two emitters involved require totally different excitation conditions. Given the ubiquity of the ERE in a wide range of interstellar environments \citep{wv04}, it is not likely that the concentration of the ERE in the walls of the RR outflow cavity is the result of an abundance enhancement but rather reflects the specific excitation requirement of the ERE process. As reviewed by \citep{wv04}, there are several independent indicators suggesting that photon with energies E $>$ 7.25 eV are required to initiate the ERE.  These energies are much higher than those required for the excitation of fluorescence in neutral PAH molecules and the number of such high-energy photons are much more limited in the SED of HD 44179. Thus, they can barely penetrate the walls of the outflow cavity before their number is exhausted, thus limiting the spatial extent of  the ERE-producing emitters. The topic of the ERE excitation will be the subject of another publication by the present authors.

\subsection{Attenuation of HD 44179}
The attenuation curve of HD 44179 appears highly unusual when compared to the wavelength dependence of interstellar extinction, but it appears to be fully consistent both qualitatively and quantitatively with expectations, if the scattered light of HD 44179 has to penetrate a large column density (A$_\mathrm{V}$ = 4.2 mag) of circumstellar material filled with neutral PAH molecules with atomic masses less than 250 a.m.u.

With a column density of carbon in the form of small PAH molecules estimated from the strength of the far-UV absorption related to PAH ionization, we can estimate a molecular abundance of such PAH molecules with typically 16 carbon atoms per molecule relative to hydrogen in the RR of about $10^{-5}$. This agrees well with the canonical abundances of PAH molecules derived from the analysis of the strength of the mid-IR aromatic emission features, commonly referred to as the unidentified infrared emission bands \citep{ld87}. This makes PAH molecules the most abundant species of interstellar molecules after H$_2$ and CO. The fact that despite the strong indications for a presence of small PAH molecules there is no clear counterpart for the interstellar 217.5 nm absorption feature in the attenuation curve of HD 44179, suggests that proposals to associate the 217.5 nm feature with PAH absorption in current dust models should be viewed with caution. The current observations appear to prohibit at least a role for small PAHs with masses $<$ 250 a.m.u. This still leaves the possibility that larger PAH structures could be involved in contributing to the interstellar 217.5 nm feature, because they would have the advantage of greater thermodynamic stability under interstellar conditions. The small PAH molecules apparent in the RR are able to form and survive in the relatively benign radiation environment of the RR, especially in the shadow of the optically thick dust ring blocking direct radiation from the central source. It is in these parts of the RR nebula where the BL intensity is strongly enhanced in comparison to the scattered radiation and the ERE. These small PAH molecules, however,  are not likely to survive the harsher radiation conditions prevailing in the diffuse interstellar medium \citep{jochims94,jbl99}, where they would be subject to photo-dissociation. As a consequence, it does not appear likely that BL with spectral characteristics and intensities similar to those observed in the RR will be seen in dusty star-forming regions such as reflection nebulae and H\textsc{ ii} regions \citep{rw75}.

The conditions deduced for the RR environment also places interesting constraints on the carriers of the diffuse interstellar bands (DIBs). Current ideas for the carriers of these much-studied but still unidentified interstellar absorption features concentrate on carbonaceous molecules, their ions in particular \citep[see][for a review]{snow95}. PAH ions have been considered as the most likely carriers by a number of investigators. Given the strong indicators for a large column density of PAH molecules and their ions in the RR, it is remarkable that only one DIB $\lambda$5780 is detected in the spectrum of HD 44179 \citep{hobbs04}. \citet{snow04} has proposed that this fact might be explained, if DIBs are produced by PAH cations with masses well in excess of 250 a.m.u. Another fact that must be taken into consideration is the absence of absorptions resulting from the D$_2\leftarrow$D$_0$ transition of anthracene$^+$ (708.76$\pm$0.13 nm), naphthalene$^+$ (670.73$\pm$0.06 nm) \citep{sukhorukov04}, and acenaphthene$^+$ (646.3 nm)\citep{biennier03} which might be expected in the spectrum of HD 44179, if these cations are present in the light path. Some of these discrepancies could be resolved by considering the ionization of these PAHs to the dication stage.

\section{CONCLUSIONS}
We summarize our conclusions:
\begin{enumerate}
\item Our spatially resolved observations of the BL in the RR have yielded the first tentative evidence that the detailed spectrum of the BL is spatially variable, consistent with a change in the dominant size of the fluorescing molecules with position relative to the central source.
\item The spatial distribution of the band-integrated BL in the RR differs fundamentally from those of the dust-scattered radiation and of the ERE, consistent with an origin from a nebular volume located mainly outside the outflow cavity and concentrated in the shadow of the disk obscuring the central source. This supports our earlier conclusion that the BL is produced by fluorescence from small, neutral PAH molecules.
\item There exists an excellent spatial correlation between the BL and the distribution of 3.3 $\micron$ C--H stretch emission from PAHs. The latter emission is predominantly produced by small, neutral PAH molecules, and the close spatial correlation supports the suggestion that the BL has its origin in the similar molecules.
\item By combining our observations with existing IUE data for HD 44179, we have determined the UV/optical attenuation curve for the central source in the RR. The attenuation curve is characterized by an exceptionally high value of R$_\mathrm{V}$ = A$_\mathrm{V}$/E(\bv) = 11.9, a remarkable  absence of the 217.5 nm absorption bump, the latter being a characteristic of Galactic interstellar extinction, and a strong, discontinuous rise in attenuation near $\lambda^{-1}\approx$ 6.0 $\micron^{-1}$. These unusual characteristics suggest strongly that the reddening observed in the light from HD 44179 is a result of radiative transfer in the RR and not due to interstellar extinction.
\item The far-UV rise in the attenuation curve of HD 44179 is qualitatively and quantitatively different from the much more gradual far-UV rise in Galactic extinction curves. The rise is, however, consistent with the onset of photo-ionization of small PAH molecules with masses of less than 250 a.m.u., if the abundance of such PAH molecules relative to hydrogen is about $10^{-5}$.  This represents a fully independent confirmation of the presence of small PAH molecules in the RR environment, consistent with our conclusion regarding the origin of the BL in the RR.
\item The broad mid-UV hump in the attenuation curve of HD 44179 whose appearance is quite unlike that of the familiar 217.5 nm interstellar absorption band, can be explained satisfactorily as resulting from a superposition of the (typically) four mid-UV absorption bands characteristic of neutral PAH molecules. An observable consequence of this absorption and the resulting electronic excitation is de-excitation via fluorescence, which we indeed observe.
\item We do not expect that BL with comparable relative intensities as seen in the RR will be seen in star-forming regions such as reflection nebulae and HII regions, where the mid-IR UIR band emissions are otherwise strong. The small PAH molecules required by the BL are not expected to survive in the radiation fields typical of star-forming regions.
\end{enumerate}

\acknowledgments
We thank David Malin and Hans van Winckel for providing us with blue images of the Red Rectangle, Laurent Verstraete and W. W. Jochims for supplying us with data on ionization cross sections of PAH molecules, and Lewis Hobbs, Theodore Snow and Donald York for constructive discussions about the Red Rectangle nebula and its central source. Some of the data presented in this paper was obtained from the Multimission Archive at the Space Telescope Science Institute (MAST). STScI is operated by the Association of Universities for Research in Astronomy, Inc., under NASA contract NAS5-26555. Support for MAST for non-HST data is provided by the NASA Office of Space Science via grant NAG5-7584 and by other grants and contracts. UPV acknowledges a CTIO thesis student travel grant. Financial suport for this study was provided through NSF Grant AST0307307 to The University of Toledo. This research has made use of NASA's Astrophysics Data System (ADS) Bibliographic Services and the SIMBAD database, operated at CDS, Strasbourg, France.


\appendix
\section{Determination of BL Intensity at the Balmer Jump}

Assume that the ratio of the scattered light intensity at $\lambda$ less than the Balmer jump (BJ) to the scattered light intensity at $\lambda$ greater than the BJ, $I_{<\mathrm{BJ}}^{sc\star}/I_{>\mathrm{BJ}}^{sc\star}$ is determined by the stellar spectrum. We measure $\lambda_{<\mathrm{BJ}}$ at 357 nm and $\lambda_{>\mathrm{BJ}}$ at 397 nm. Assuming a constant albedo over the wavelength range and an optically thin scenario, the relative increase in the scattered flux at 357 nm compared to that at 397 nm is 1.08 \citep{savage79}. This factor is an upper limit as it will be smaller in an optically thick case. 
Thus, in the stellar spectrum
\begin{displaymath}
\frac{I_{357}^{sc\star}}{I_{397}^{sc\star}}=0.2334
\end{displaymath}
After correction for wavelength dependent scattering,
\begin{displaymath}
\frac{I_{357}^{sc\star}}{I_{397}^{sc\star}}=0.2520
\end{displaymath}
$\Longrightarrow$
\begin{displaymath}
I_{357}^{scN}=0.2520*I_{397}^{scN}
\end{displaymath}
where the superscript $N$ denotes the values for the nebular spectra. Also,
\begin{eqnarray*}
I_{357}^{totalN}&=&I_{357}^{BL}+I_{357}^{scN}\\
&=&I_{357}^{BL}+I_{397}^{scN} * 0.2520
\end{eqnarray*}
Therefore, the intensity of the BL at 357 nm, \\
\begin{displaymath}
I_{357}^{BL}=I_{357}^{totalN} - I_{397}^{scN} * 0.2520
\end{displaymath}
which can be determined by knowing the value of the scattered light intensity at $397$ nm, which in turn is previously determined by subtracting the BL intensity (measured using the line-depth technique) at that wavelength from the total intensity at that wavelength.

\end{document}